\documentclass[twocolumn,showpacs,preprintnumbers,amsmath,amssymb]{revtex4}
\usepackage{graphicx}
\usepackage{dcolumn}
\usepackage{bm}
\usepackage{epsfig}
\newcommand{\beq}{\begin{equation}}
\newcommand{\eeq}{\vspace{0cm} \end{equation}}
\newcommand{\beqq}{\setlength\arraycolsep{2pt}
\begin{eqnarray}}
\newcommand{\eeqq}{\vspace{0cm} \end{eqnarray}}

\begin{document}
\title{From de Sitter to de Sitter: A Thermal Approach to Running Vacuum Cosmology and the Non-Canonical Scalar Field Description}
\author{P. E. M. Almeida$^a$\footnote{pedro.emalmeida@usp.br}}
\author{R. C. Santos $^{b}$\footnote{rose.clivia@unifesp.br}}
\author{J. A. S. Lima$^{a}$\footnote{jas.lima@iag.usp.br}}
\affiliation{$^{1}$Departamento de Astronomia, Universidade de S\~{a}o
Paulo \\ Rua do Mat\~ao, 1226 - 05508-900, S\~ao Paulo, SP, Brazil}
\affiliation{$^b$Universidade Federal de S\~ao Paulo (UNIFESP) Diadema, SP, Brazil} 

\pacs{98.80.-k, 95.36.+x}

\bigskip
\begin{abstract}
 \noindent The entire classical cosmological history between two extreme de Sitter vacuum solutions is discussed based on Einstein's equations and non-equilibrium thermodynamics. The initial non-singular de Sitter state is characterised by a very high energy scale which is equal or smaller than the reduced Planck mass.  It is structurally unstable and  all the continuous created matter, energy and entropy  of the material component comes from the irreversible flow powered by the primeval vacuum energy density. The analytical expression describing the running vacuum is obtained from the thermal approach. It opens a new perspective to solve the old puzzles and current observational challenges plaguing the cosmic concordance model driven by a rigid vacuum.  Such a scenario is also modelled through a non-canonical scalar field. It is demonstrated that the resulting scalar field model is shown to be step by step a faithful analytical representation of the thermal running vacuum cosmology.
\end{abstract}

\maketitle

\section{Introduction} Since long ago, investigations on de Sitter and quasi-de Sitter cosmic solutions have attracted much attention along different research lines.  Such studies were somewhat inspired by  the possible avoidance of the primeval singular state ordinarily predicted by\,``big-bang" models,  the decaying vacuum emerging from a plethora of inflationary models driven by scalar fields, including or not quadratic corrections to Einstein gravity \cite{G1966,S1966,M1973,S80,A83, A84,Davies82}. A different approach analysed the possible creation of the universe from ``nothing" (in the sense of Vilenkin \cite{AV1985}), with the classical universe emerging as a de Sitter solution from an expected but still observationally unavailable quantum gravity regime. This primeval state without ordinary matter may be characterised by a physical energy scale, $H_I \leq M_P$, or equivalently, by  a vacuum energy density, $\rho_I=3M_P^{2}H_I^{2}$, where   $M_P = ({8\pi\,G})^{-1/2} \simeq 2.4\times 10^{18}$ GeV is the reduced Planck mass (in our units $\hbar=k_B=c=1$).

It is also widely known that a non-singular {\it de Sitter spacetime is structurally unstable} both from quantum and classical viewpoints, at least due to three different effects:  Quantum corrections on the geometric sector of general relativity\,\cite{S80}, the existence of thermal fluctuations \cite{EM86}, and the process of gravitational particle creation in the expanding universe \cite{Davies82,PRI89,PRI89_2,CLW92,CLW92_2,LG92},  Henceforth, due to its simplicity and interesting physical consequences for the evolution of the emergent classical initial vacuum state, the presence of thermal instabilities conjoined with an irreversible flow of energy, matter and entropy forming the thermal bath will be taken for granted. Additional reasons will be outlined ahead. 

 More recently, independent astronomical observations have shown that the cosmic concordance model ($\Lambda$CDM + Inflation) provides a quite reasonable and predictive description of the current universe driven  by a rigid vacuum energy density, $\rho_{\Lambda_0} = M^2_p\Lambda_0$. However, the incredible discrepancy with what is expected from quantum field theory (QFT) as compared to current astronomical observations, $\rho_{\Lambda_0}/\rho_{\Lambda_I} \sim 10^{-122}$, gave rise to the cosmological constant problem (CCP)\,\cite{SW89}. Furthermore, according to the cosmic concordance model the dominance of rigid vacuum, ($\rho_{\Lambda_0} \simeq \rho_M$), where $\rho_M$ is the matter density, took place  at a redshift $z \simeq 0.55$ (see e.g., \cite{V2014}) thereby leading to the so-called coincidence problem (CP). Another well known consequence of the rigid vacuum $\Lambda$CDM cosmology is the inexorable evolution of the current Universe to a final de Sitter spacetime. Here it is characterised by an extremely low vacuum energy density, say,  $\rho_{\Lambda_F} = 3 {M_P^2}{H_F^2}$, where the Hubble parameter $H_F$ provides the  final de Sitter constant energy scale, which will be attained only for $\rho_M\equiv0$.
 
 Currently, the standard rigid vacuum cosmology is also being observationally challenged by the $H_0$ and $S_8$ tensions. In the case of $H_0$, local determinations of the Hubble constant ($H_0$) using Cepheid-calibrated supernovae of the SHOES collaboration \cite{R2019}, measurements at intermediate redshifts from a combination of tests \cite{LCM2007, LCM2007_2} and strong lensing time-delays \cite{B2019}, which in comparison with the independent prediction of Planck collaboration plus $\Lambda$CDM, shows a clear-cut tension around 5$\sigma$ \cite{P2018}. In addition, estimates based on cosmic shear evaluations from  weak lensing are favouring values of the parameter $S_8=\sigma_8\sqrt{\Omega_M/0.3}$, where $\sigma_8$ measures the current mass fluctuation in $8 h^{-1}$Mpc, which are lower than the ones provided by early times probes \cite{kids2020,Val2021,ML2021}. 

 In this context, it seems natural to advocate that the whole classical evolution of the Universe would be analytically described between two de Sitter stages defined by a pair of extreme and constant energy scales $H_I$ and $H_F$. In this case, for all theoretical purposes, we may also consider that  the second de Sitter vacuum state of very low energy density is the true vacuum state. In certain sense,  the structural instability of the primeval de Sitter ($H_I$) is somewhat driving the evolution of the observed universe to its `ground state', corresponding to the final de Sitter vacuum, $H_F < H_0$, where the sub-index 0 refers to present day quantities. Thus, it is not surprising that a lot of attention has been paid to running vacuum cosmologies in the last decade by several reasons, among them: ({\bf i}) A simple solution to the singularity problem, as well as, the old CCP and CP puzzles, ({\bf ii}) a complete cosmology ({\it from de Sitter to de Sitter}) with two accelerating stages driven by the same actor, described by  the running vacuum  energy density,  $\rho_{\Lambda}(H_I,H_F,H)$, departing slightly at all phases from the rigid vacuum model plus slow-roll inflation, and, ({\bf iii}) a possible solution to the current $H_0$ and $S_8$ observational tensions plaguing the current standard cosmology  provided by the $\Lambda$CDM model, which at late times model is driven by a rigid vacuum \cite{ST2009,LBS2013,PLBS2013,PLBS2013_2,J112015,GV2017,S2017,ZSL2018}. 

As we shall see, a suitable $\Lambda(H)$-term uniting both de Sitter phases can be deduced using non-equilibrium thermodynamics and the Einstein field equations ({\bf EFE}). The classical evolution of the Universe can also be analytically described and some of the questions outlined above are answered. Perhaps more interesting, we demonstrate that a minimally coupled ordinary scalar field cannot describe the irreversible two-fluid approach of the running vacuum scenario. However, the whole evolution can  perfectly be mimicked by a single minimally coupled non-canonical field when interpreted as a mixture of two minimally coupled interacting perfect fluids  \cite{OT86,OT86_2,KF1987,CW1990,AR92,JCW92,IW93,IW93_2,L96, L96_2}.

The paper is organised as follows. In section II we set up the Einstein equations for an interacting mixture of running vacuum medium plus a perfect fluid. In section III, the irreversible thermodynamic approach for such a mixture is discussed in detail. In particular, by taking into account only the thermal effects of the interacting mixture, a viable expression for $\Lambda$(H) is phenomenological deduced based on Einstein's equations and non-equilibrium thermodynamics. The equation governing the entire classical evolution of the Universe from $H_I$ to $H_F$ is also determined. Section IV is dedicated to the basic results of the cosmic eras and the solutions of some cosmological problems are also discussed. In section V, a representation of the whole process in terms of a non-canonical scalar field and the associated dynamics is investigated. The article is closed in section VI with the final comments and conclusions. In the Appendix C, the two basic parameters of the non-canonical scalar field potential are calculated in terms of $H_I$, $H_F$, $M_P$ and the mass scale ($M$) of the scalar field.
\section{Basic Equations}

In what follows we consider a class of Friedmann-Lema\^itre-Roberson-Walker (FLRW) cosmologies described by a flat geometry:
\begin{equation}
 ds^{2} =dt^{2} - a^{2}(t)\left(dx^{2} + dy^{2} + dz^{2}\right), 
\end{equation}
where $a(t)$ is the scale factor.
In such a background, the {\bf EFE} for a perfect fluid plus a running vacuum can be written as:
\begin{eqnarray}\label{EFE}
	\rho\,+\,M_P^{2}\Lambda(H) &=& 3M_{p}^2\,H^2\;,\label{eq1}\\
   p - M_P^{2}\Lambda(H) &=& -M_{p}^2\left[2\dot H + 3H^{2}\right]\,,\label{eq2}
	\end{eqnarray}
where $\rho$ and $p$ are the energy density and pressure of the ordinary fluid component while a dot means comoving time derivative. Now, by adopting the ``$\omega$-law" equation of state (EoS):
\begin{equation}\label{EoS}
p=\omega \rho, \,\,\,\,\, {\bf  0 \leq \omega \leq  1}, 
\end{equation}
it is readily seen that equations (\ref{eq1})-(\ref{EoS}) imply that the cosmic dynamics is driven by the ``H-equation of motion" 
\begin{equation}\label{Eqm1}
 \dot H + \frac{3(1+\omega)}{2}H^2 \left[1 - \Omega_{\Lambda}(H)\right]=0\,, 
\end{equation}
where $\Omega_{\Lambda} (H)=\Lambda(H)/3H^{2}$ is the vacuum density parameter. 

In principle, for $\Omega_{\Lambda}=1$,  it is possible to have two extreme de Sitter solutions ($\dot H = 0$) describing the initial and final vacuum states, say, $H=H_I$ and $H=H_F$, and, as such, two accelerating cosmic states (early and late time inflation) may result from the same actor \cite{LBS2013, PLBS2013,ZSL2018}.  The initial de Sitter spacetime ($H_I$) works like a “repeller” (unstable solution) at early times. However, due to the cosmic evolution, energy, particles and entropy are being continuously transferred to the fluid component thereby giving rise to an attractor de Sitter spacetime in the distant future, characterised by a very low energy scale ($H_F$). Accordingly, $\Omega_{\Lambda}=1$ implies that the fluid energy densities for both extreme scales are nullified, that is, $\rho (H_I)=\rho (H_F) = 0$. 

Of course, the complete evolution driven by (\ref{Eqm1}), linked to the subsequent radiation-vacuum and matter-vacuum dominated phases, comes out only if the expression for $\Lambda(H)$ is known. Next, it will be derived by combining the {\bf EFE}, non-equilibrium thermodynamics and the limiting `boundary' constraints provided by the extreme de Sitter solutions. 

\section{Running Vacuum: A thermal Approach}
Let us now consider the energy flow from the decaying vacuum creating particles and entropy to the fluid component. Following standard lines, the thermodynamic states for the interacting mixture is defined by the energy conservation law ($u_{\mu} T^{\mu\nu}_{;\nu}= 0$) and the balance equations for the number of particles and entropy fluxes, $N^{\mu} = nu^{\mu}$ and $S^{\mu}=su^{\mu}$: 
\begin{eqnarray}\label{TEQ}
u_{\mu} T^{\mu\nu}_{;\nu}&=& 0 \,\,\,\, \Longleftrightarrow \,\,\,\,\ \dot{\rho} + 3H(1 + \omega)\rho= -M_{P}^2\dot{\Lambda}, \\ \label{n_eq}
N^{\mu}_{;\mu} &=& n\Gamma \,\,\,\, \Longleftrightarrow \,\,\,\,\ \dot n + 3nH = n\Gamma,
\label{n_gamma}
\\ 
S^{\mu}_{;\mu} &=& s\Gamma \,\,\,\, \Longleftrightarrow \,\,\,\,\ \dot s + 3sH = s\Gamma,
\label{s_gamma}
\end{eqnarray}
where $n$ and $s$ are the particle concentration and entropy density and $\Gamma$ is the particle creation rate, being the same for particle number and entropy. Thus, the entropy of the fluid is also a pure consequence of the decaying vacuum, and, as such, the bulk viscosity process and gravitational matter creation have been neglected\,\cite{LS2021}. 
In principle, one may argue that the 2nd law of thermodynamics should be applied for both components. However, the vacuum fluid behaves like a condensate carrying no entropy, as happens in the two-fluid description ordinarily employed in superfluid dynamics\,\cite{LL86}. The vacuum state is assumed to have energy but no entropy and real particles. 

In the course of the decaying vacuum process, one may see from (7)-(8) that the specific entropy, $\sigma=s/n=S/N$, remains constant ($\dot\sigma = 0$). Such  an interesting result becomes more clear when rewritten as (see Appendix A): 
\begin{equation}\label{dotS}
\frac{\dot S}{S} = \frac{\dot N}{N} = \Gamma \geq 0.
\end{equation}
This inequality implies that the decaying vacuum (due to its own energy density) can only produce matter in the space-time ($\dot N \geq 0$), while the reverse process is thermodynamically forbidden. 

The condition $\dot \sigma =0$ has another remarkable consequence. Actually, by combining Gibbs' law, $nTd\sigma= d\rho - (\rho + p) dn/n$, with (4) and (5), we find a key result in this framework:
\begin{equation}\label{dotL}
\dot \rho - (\rho + p)\frac{\dot n}{n} = 0\,\, \leftrightarrow\,\,  M_P^{2}\dot\Lambda= -(\rho + p)\Gamma.
\end{equation}
The second equality above is a typical first-order thermodynamic relation uniting the stress $\dot \Lambda$ (``flux") to its thermodynamic force $\Gamma(H)$. The above coefficient of $\Gamma$ is the fluid enthalpy density which due to the weak energy condition satisfies the inequality $h\equiv\rho+p = (1+\omega)\rho \geq 0$.  

Notice that  by adding {\bf EFE} (\ref{eq1})-(\ref{eq2}) we find $\rho + p = - 2M_{p}^2 \dot H$, which means that $\dot H < 0$. Thus, since $\Gamma\geq 0$ from (\ref{dotS}), it follows that the constraint $\dot\Lambda \leq 0$ must be obeyed. In this thermal approach, a decaying vacuum from the primeval de Sitter state is also required  by the second law of thermodynamics. 

Now, by inserting $h$ into (\ref{dotL}), a new and interesting result is derived:
\begin{equation}\label{Leq}
\Lambda(H) = 2\int\Gamma(H)dH + B,  
\end{equation}
where $B$ is a constant to be fixed by the de Sitter `boundary conditions'. To the best of our knowledge, the above integral expression defining  $\Lambda$(H) in terms of the creation rate $\Gamma(H)$, has not been found before. Thus, in order to obtain a complete description (from $H_I$ to $H_F$),  an expression for $\Gamma(H)$ need to be  specified.

At this point, it should be remarked that until now only classical macroscopic tools were explored, namely: {\bf EFE} and non-equilibrium thermodynamics. It is also well known that particle creation from a perturbed vacuum state is essentially a microscopic phenomenon. Hence, from a more rigorous viewpoint, an expression for $\Gamma(H)$ in this context cannot be macroscopically deduced. Nevertheless, under the proviso that the dimension of $[\Gamma]=[H]$, a simple possibility would be  $\Gamma (H) = 3\nu H$, where the factor 3 is introduced for mathematical convenience and $\nu$ is a dimensionless positive free parameter, in principle, restricted on the interval $0 \leq \nu < 1$.   

Furthermore, such a linear expression  generates a term $\Lambda (H) \propto H^{2}$. However, it is not enough for describing the unified cosmic history from de Sitter to de Sitter since it generates a singular initial state \cite{JCW92}. At late times, due to the very low expansion rate, $\nu$ is taken as a constant parameter. However, it may be a function of $H$ at early times thereby implying a more intense transference of energy, particles and entropy to the material component. This is needed  to the formation of the primeval thermal bath both in the macroscopic approach \cite{LBS2013}, as well as for a possible scalar field description \cite{LA2021}. 

In this connection, a renormalized vacuum energy density proportional to $H^{4}$ has been derived long ago  \cite{Davies82}. This is also the next term obtained by Shapiro and Sol\`a based on the covariance of the renormalized action \cite{SS2002}. From (\ref{Leq}) we also see that a term proportional to $H^{4}$ can also be generated by a creation rate $\Gamma(H) \propto H^{3}$. Therefore, inspired by such studies and also some ad hoc treatments for $\Lambda(H)$ \cite{ST2009,LBS2013,ZSL2018} a quadratic correction, namely,  $3\gamma(H/H_I)^{2}$, where the arbitrary $\gamma=\gamma(\nu)$, is added to the natural term based on dimensional grounds. In particular, for $\gamma=2(1-\nu)$ it follows that
\begin{equation}\label{gamma}
\Gamma(H) = 3\nu H  + 6(1-\nu)H\left({H}/{H_I}\right)^{2},  
\end{equation}
whereas $\Lambda(H)$ as given by (\ref{Leq}) reads:
\begin{equation}\label{LH}
\Lambda(H) = 3(1-\nu)H_F^{2} + 3\nu H^{2} + 3(1-\nu){H^{2}} \left(\frac{H}{H_I}\right)^{2}.
\end{equation}
As a simple check of such expressions, one may compute directly $\Gamma (H) ={d\Lambda(H)}/{2dH}$ (see Appendix B for a more detailed calculation including both de Sitter boundary conditions).  

\begin{figure*}
\includegraphics[scale = 0.5]{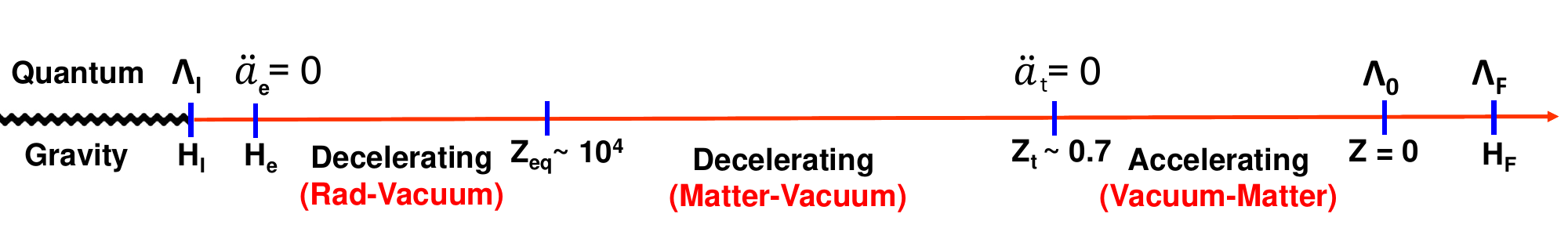}
\caption{From de Sitter to de Sitter (out of visible scale). Just after the unknown quantum gravity regime, all the energy of the Universe is concentrated on the emergent and unstable de Sitter spacetime  ($\Lambda_I=3H_{I}^{2}, \rho_{\Lambda_I} = 3M_P^{2}H_{I}^{2},\,\rho_M\equiv 0$). The classical evolution shown above is analytically described by the decaying $\Lambda(H)$ model as given by\,(\ref{LH}) which is the unique fuel of both inflationary stages. Since $\Lambda(H)$ is a continuous function, the equilibrium redshift $z_{eq}$ is determined in the usual way (sudden transition) with a small contribution of the $\nu$ parameter as discussed in (\ref{zeq}). For all allowed values of $\nu$, the ultimate destiny of the model is the pure de Sitter vacuum state fixed by  $\Lambda_F=3H_F^{2}$. However, it is attained only in the distant future ($a\rightarrow\infty$), as shown in (\ref{H_sol_2}).  For $\nu=0$, the late time evolution after ${\ddot a} (H_e) = 0$ is just the rigid vacuum $\Lambda$CDM model.} 
\end{figure*}

Now, it is easy to see  that the expressions for the density parameters take the form: 
\begin{eqnarray}\label{OMega}
\Omega_{\Lambda}(H) &=& \nu + (1-\nu)\frac{H_F^{2}}{H^{2}} + (1-\nu)\left(\frac{H}{H_I}\right)^{2}, \label{OL}\\ 
\Omega_M (H) &=&(1-\nu)\left[1-\frac{H_F^{2}}{H^{2}} - \left(\frac{H}{H_I}\right)^{2}\right], \label{OM}
\end{eqnarray}
where $\Omega_M$ is the matter density parameter. Now, by inserting $\Omega_{\Lambda}$ or $\Omega_M$  into (\ref{Eqm1}), it is readily seen that the classical evolution of the model between the two extreme de Sitter phases is governed by: 
\begin{equation}\label{eqmf}
\dot H + \frac{3(1+\omega)(1-\nu)H^2}{2} \left[1 - \frac{H_F^{2}}{H^{2}}- \frac{H^{2}}{H_I^{2}}\right]=0. 
\end{equation}
 
 The above equation depends on two pairs of positive parameters ($\omega,\,\nu$) and two energy scales ($H_I,\, H_F$). Like in the rigid vacuum model $\omega$ is differently fixed by the description of the cosmic eras $\omega=1/3$ (radiation) and $\omega=0$ (dust). So, it cannot rigorously be considered a new free parameter. For the remaining three parameters, we remark that the energy scale $H_I$ has a natural upper bound given by the reduced Planck mass $M_P$ (see introduction), while the final $H_F$ scale can be determined in order to solve the CCP problem, by assuming, for instance, that $H_I=M_P$. It can also be calculated by the model in terms of $H_0$. 
 
{\bf Figure 1} displays schematically the complete evolution of the whole classical scenario. 
 Observe that (\ref{eqmf}) remains valid even for $\nu\equiv 0$ since this minimal model is also defined by the same extreme constant physical scales ($H_I$, $H_F$). The  only difference is that the late time evolution for $H<<H_I$ happens in such a way that CMB photons are not produced anymore. Moreover, it is easy to see that the scenario becomes the standard, rigid  $\Lambda$CDM cosmology described here in terms of the constant final scale $H_F$. Thus, only the early inflationary stage is modified because it started from a pure de Sitter vacuum solution. The details of each transition describing the evolution (from de Sitter to de Sitter) through different cosmic eras will be discussed next. 
 
\section{Cosmic Eras}

\subsection{From initial pure de Sitter Vacuum to Radiation-Vacuum Phase}

In the primeval universe,  the EoS  parameter is  $\omega =1/3$, and the Hubble scale is restricted on the interval $H_I \geq H >> H_F$.  In this limit, (\ref{eqmf}) reduces to: 
\begin{equation}\label{eqmf1}
\dot H + 2(1-\nu)H^2\left[1 - \frac{H^{2}}{H_I^{2}}\right]=0,
\end{equation}
whose solution in terms of the scale factor reads: 
\begin{equation}\label{HrS}
H(a) = \frac{H_I}{[1 + Ca^{4(1 - \nu)}]^{1/2}},
\end{equation}
where $C = a_e^{-4(1 - \nu)}/(1 - 2\nu)$ is an integration  constant.  

In the same limits for $H$ given above, we see that the vacuum energy density reduces to:
\begin{equation}\label{rhoL1}
  \rho_{\Lambda}(H) = 3\nu M_P^2 H^2 + 3(1 -\nu) M_P^2 H^2\left(\frac{H}{H_I}\right)^2,
\end{equation}
and by inserting  solution (\ref{HrS}) for $H(a)$ into the above expression we obtain:
\begin{equation}\label{rhoL2} 
 \rho_{\Lambda}(a)= \rho_I\frac{1 + \nu Ca^{4(1 - \nu)}}{\left[1 + Ca^{4(1-\nu)}\right]^{2}},
\end{equation}
where $\rho_I = 3M_P^2H_I^2=\rho_{\Lambda}(H=H_I)$.

The radiation energy density can be obtained by using use the value of $\Omega_M$  or even the  Friedmann equation {\bf EFE} (\ref{eq1}), and combining the result with the above $H(a)$ solution. It follows that
\begin{eqnarray}\label{rhor}
  \rho_R (H) &=& 3(1 - \nu)M_P^2 H^2\left[1 - \left(\frac{H}{H_I}\right)^2\right],\\
  \rho_R (a) &=& \rho_I\frac{(1 - \nu) Ca^{4(1 - \nu)}}{\left[1 + Ca^{4(1-\nu)}\right]^{2}}.
\end{eqnarray}
Note that all the above equations are in agreement with the physical limits. In particular for $H=H_I$, $\rho_R(H_I)\equiv 0)$ as expected for a pure de Sitter vacuum state. In this scenario, the classical instability of the initial de Sitter state, which is analytically  described by (\ref{HrS}), is the unique source of the primeval thermal bath. 

Let us now examine with more detail when this running $\Lambda(H)$ model performed its transition from the very early inflation dominated by the vacuum state to the radiation-vacuum phase. In principle, since $\omega =1/3$, this must happen for a value of $H_e$ (see also {\bf Figure 1}), which is determined by the equality of the energy density of both components, that is,  $\rho_{\Lambda}(H_e) = \rho_R(H_e)$. From equations (\ref{rhoL1}) and (\ref{rhor}), one finds:

\begin{equation}\label{He}
H_e = \sqrt{\frac{1-2\nu}{2(1-\nu)}}H_I.
\end{equation}
 The meaning of the $H_e$ result can also be understood by taking the deceleration parameter into account. By definition 
\begin{equation} \label{dec_par}
  q(H)= - \frac{\ddot{a}a}{\dot{a}^2} = - 1 - \frac{\dot{H}}{H^2} = -1 + 2(1-\nu)\left[1 - \frac{H^{2}}{H_I^{2}}\right].
\end{equation}
where in the last equality we have used the `H-equation of motion'. For ending inflation (${\ddot a=0}$) at $H=H_e$, the above equation yields,  $H_e = [(1-2\nu)/2(1-\nu)]^{1/2}H_I$, which is exactly the same value determined by the densities equality condition.  Therefore, different from many adiabatic scalar field models, the value of $H_e$ provides both the end of inflation and the beginning of the radiation dominated phase \footnote{Such a result is not surprising because from {\bf EFE} one finds $\ddot a= -(6M_p^2)^{-1}[(1 + 3\omega)\rho - 2\rho_{\Lambda}]a$. Thus for $\omega=1/3$, $\ddot a = 0$ implies $\rho\equiv \rho_R =\rho_{\Lambda}$ while for $\omega=0$, one finds  $\rho\equiv \rho_M = 2\rho_{\Lambda}$.}.

It should also be remarked that some observations constrain the values of $\nu$ to be much smaller than unity \cite{BS2009}. This is an advantage of such models since the evolution after entering the radiation-vacuum phase is not too drastically different from $\Lambda$CDM. The same occurs at low redshifts when $\ddot a(z_t)=0$. In this connection, recent studies have also pointed out that a possible explanation to the current $H_0$ and $\sigma_8$ tensions require mildly deviations of $\Lambda$CDM both at early and late times\,\cite{GV2017,Val2021,ML2021}. 

\subsection{From Radiation-Vacuum to Matter-Vacuum Dominated Phase}

In this limit we assume $\omega=0$ and $H_F \leq H \ll H_I$ so that (\ref{eqmf}) now boils down to:

\begin{equation}\label{eqmf2}
\dot H + \frac{3(1-\nu)H^2}{2} \left[1 - \frac{H_F^{2}}{H^{2}}\right]=0, 
\end{equation}
whose solution is given by: 

\begin{equation}\label{H_sol_2}
  H(a) = H_F[1 + Da^{-3(1 - \nu)}]^{1/2} = H_F[1 + D(1+z)^{3(1 - \nu)}]^{1/2}, 
\end{equation}
where the integration constant $D$ is readily obtained computing the current value of (\ref{H_sol_2}). For the sake of simplicity, we fix  $a_0 \equiv 1$ and defining $H(a_0) \equiv H_0$,  it follows that $D = (H_0/H_F)^{2} - 1$. The above expression also shows that the ultimate value of the Hubble parameter ($H_F$)  is attained only in the distant future, $a \rightarrow \infty$, or equivalently, $z \rightarrow -1$.

Note also that in the same limit, from (\ref{LH}) and the first {\bf EFE}, the vacuum and matter (dust) energy densities are easily obtained: 
\begin{eqnarray}
  \rho_{\Lambda}(H) &=& 3(1 - \nu)M_P^2H_F^2 + 3\nu M_P^2 H^2, \label{rhoL} \\
\rho_M(H) &=& 3(1 - \nu)M_P^2 H^2\left[1 - \left(\frac{H_F}{H}\right)^2\right]. \label{rhoM} 
\end{eqnarray}
Now, let us determine the transition redshift  which generalises the time of radiation-matter equivalence of the rigid vacuum model. Here we take it to be $t=t_e$ which is akin to $z(t_{eq})= z_{eq}$. We will derive this result from the evolution of the temperature law in the presence of a decaying $\Lambda(H)$ which for $H<<H_I$ and $\Gamma = 3\nu H$ is given by \cite{L96}

\begin{equation}\label{TL1}
    T = T_0(1 + z)^{1- \nu}.
\end{equation}
At this time the initial vacuum ($H_I$) is almost completely depleted since the transition radiation-matter should occur at the end of the radiation phase. In addition, since $\dot {\sigma} = 0$ for an\,`adiabatic' decaying vacuum, the equilibrium relations of the CMB are preserved, that is, $\rho_R \propto T^{4}$ and $n_R \propto T^{3}$. This means that the radiation temperature is given by:
\begin{equation}
    \rho_R = \rho_{R0} \left(\frac{T}{T_0}\right)^{4} = \rho_{R0} (1+z)^{4(1-\nu)} 
\end{equation}
where $\rho_{R0}$ is the present day radiation energy density.
Substituting the temperature law in (\ref{TL1}) and taking the equality between radiation and matter density we  obtain the transition redshift $z_{eq}$:

\begin{eqnarray}
    \rho_{R0}(1 + z_{eq})^{4(1 - \nu)} = \rho_{M0}(1 + z_{eq})^{3(1 - \nu)},
    \\
    \Rightarrow z_{eq} \simeq \left(\frac{\Omega_{M0}}{\Omega_{R0}}\right)^{1/(1-\nu)}, \label{zeq}
\end{eqnarray}
where in the approximation we have used that $z_{eq} >>1$. For 
 $\nu = 0$ and inserting the observational values, $\Omega_{M0} \approx 0.3$ and $\Omega_{R0} \sim 10^{-5}$, we obtain $z_{eq} \approx 3\times10^{4}$, which is the same result obtained in the case of the rigid vacuum,  $\Lambda$CDM model. Hence, there is a small correction for $\nu \neq 0$ whose value may be determined by constraining $\nu$ from current observations.
 
At this point, it is  interesting to know when the beginning of the second accelerating stage of the universe in the dust-vacuum phase happens, that is,  when ${\ddot a}_t=0$ (see Fig. 1). The  deceleration parameter now reads:

\begin{equation}
q(H) = - \frac{a\ddot a}{{\dot a}^{2}}= -1 + \frac{3(1 - \nu)}{2}\left[1 - \frac{H^2_F}{H^2}\right],
\label{qm}
\end{equation}
so that $q(H)=0$ leads to
\begin{equation}\label{H_2star}
H({\ddot a}_t=0) = H_F\left[\frac{3(1 -\nu)}{1 - 3\nu}\right]^{1/2}.
\end{equation}
Note also that the scale $H_F$ can be obtained in terms of $H_0$ from the expression of $\Omega_M$ as given by (\ref{OM}). For $H<<H_I$ we find
\begin{equation}\label{HF}
H_F = H_0\sqrt{\frac{\Omega_{\Lambda_0} - \nu}{1-\nu}},
\end{equation}
where the value of $H_F$ in (\ref{HF}) has been used. For small values of $\nu$, it follows that the final de Sitter is nearly characterised by $H_F \simeq \sqrt{\Omega_{\Lambda_0}}H_0 \sim 0.83H_0$.  

In the matter-vacuum dominated era, it is also convenient to rewrite the expression of the Hubble parameter (\ref{H_sol_2}) in terms of the current observable quantities. One finds
\begin{equation}
  H(\Omega_{M0}, z) = H_0\left[\frac{\Omega_{M0}}{1 - \nu}(1 + z)^{3(1 -\nu)} + 1 - \frac{\Omega_{M0}}{1 - \nu}\right]^{1/2}.
\end{equation}
Note that in the limit $z\rightarrow -1$, the same value of (\ref{HF}) obtained by a different way is recovered.
The above equation resembles the expression for the Hubble parameter in the rigid $\Lambda_0 CDM$ model, becoming exactly the same  for $\nu = 0$. In addition, by defining $\Omega_{M0}^{eff} = \Omega_{M0}/(1 - \nu)$ the above equation becomes 
\begin{equation}
  H(\Omega_{M0}, z) = H_0\left[\Omega_{M0}^{eff}(1 + z)^{3(1 -\nu)} + 1 - \Omega_{M0}^{eff}\right]^{1/2}.
\end{equation}

\subsection{Solving Some Cosmological Problems}

Historically, the  `Big-Bang' singularity predicted by the classical general relativity, has been considered one of the most challenging cosmological problems. Since long ago,  some authors suggested that a  natural classical solution in the general relativistic domain comes out whether the universe starts like a de Sitter spacetime and evolves to a radiation dominated phase (see also e.g., \cite{M1973}), similarly to what is proposed here. However, different from  several inflationary solutions, the present scenario does not need a super-cooling process followed by an extreme reheating mechanism, in order to solve the horizon and flatness (entropy) problems. In particular, the so-called\, `graceful exit' problem is absent (in this connection see also \cite{LA2021}).

The cosmic evolution in the long run ($H_I$ to $H_F$), also suggests an interesting perspective to the some cosmological constant problem. Particularly, the ratio between the extreme vacuum energy densities reads:        
\begin{equation}\label{CC}
\mathcal{R} = \frac{\rho_{\Lambda F}}{\rho_{\Lambda I}} \equiv \frac{\Lambda_{F}}{\Lambda_{I}}=\frac{H_F^{2}}{H_I^{2}}.
\end{equation}
Hence, the $\Lambda$ - problem is now reduced to the ratio between two constant scales. From the above calculated result for $H_F$ and assuming two different values for the initial scale, say,  $H_I \sim M_P$ and $H_I \sim 10^{16}$\,Gev associated to the Grand Unified Theory (GUT) scale, the values of the above ratio are ${\mathcal{R}}_{(1)} = 10^{-122}$ and ${\mathcal{R}}_{(2)} = 10^{-119}$, respectively. Such results are in agreement with naive expectations from quantum field theory. However, as remarked before, the correct value of the inflationary scale still needs an accurate observational determination from the B-mode power spectrum of CMB \cite{KAM2016}. 

One may also advocate that the coincidence problem is also naturally solved in our scenario. Its naturalness arises because $\rho_v \simeq \rho$ for two different phases. Actually, at the end of the first inflationary phase when $H=H_e$ it is easily checked that $\rho_{\Lambda}=\rho_{R}$. Moreover,  deep in the matter dominated phase, the second accelerating stage vacuum started when $\ddot a=0$ with $\rho_{\Lambda} = 2\rho_M$. Hence, such `coincidences' can also be seen  as a kind of \textit{necessity}, that is, a natural consequence of an evolution between two extreme de Sitter solutions (i.e., the same actor) constrained by two definite scales. 

It is also interesting to investigate whether this classical non-singular cosmology (from de Sitter to Sitter) driven by a running vacuum component can also be described in terms of a scalar field. This point will be investigated in detail next section. 

\section{Running Vacuum Cosmology: A Non-Canonical Scalar Field Description}

Let us now consider a pure scalar field $\phi$ minimally coupled with gravity. In the framework of general relativity the total action for a non-canonical field can be written as 
\begin{eqnarray}\label{Action} 
S = S_g + S_{\phi} = \int d^4x\sqrt{-g}\left[-\frac{M_{p}^2}{2}R +
{\cal{L}}\left(X,\phi\right) \right], 
\end{eqnarray} 
where ${\cal{L}}\left(X,\phi\right)$ is a functional of the standard kinetic term $X=\frac{1}{2}\partial^{\mu}\phi\partial_{\mu}\phi \equiv {\dot \phi}^{2}/2$. 

In what follows it will be assumed that the dominant energy contribution in the primeval Universe comes from the non-canonical scalar field, which is described by the Lagrangian \cite{VS2012} 
\begin{equation}
\mathcal{L}(X,\phi) \equiv X\left(\frac{X}{M^{4}}\right)^{\beta-1} - V(\phi), 
\label{L1}
\end{equation}
where the scale $M$ has dimension of mass, $V(\phi)$ is the potential and $\beta$ a real positive number. The energy-momentum tensor (EMT), $T^{\mu}_{\nu} =(\frac{\partial \mathcal{L}}{\partial X}) \partial^{\mu}\phi \partial_{\nu}\phi$ - $\delta^{\mu}_{\nu}{\mathcal{L}}$, \, is diagonal with components:
\begin{eqnarray}
\rho_{\phi}&\equiv&\rho_k + V(\phi)=(2\beta-1)X\left(\frac{X}{M^4}\right)^{\beta -1} + V(\phi),\,\label{ED}\\ 
 p_{\phi}&\equiv & p_k - V(\phi) = X\left(\frac{X}{M^{4}}\right)^{\beta-1}-V(\phi),\,\,\label{p}
\end{eqnarray}
where $\rho_k$ is a non-canonical kinetic term and a dot means time derivative. It is also easy to show that the following equation of motion drives the evolution of the non-canonical scalar field: 

\begin{eqnarray}
    \ddot{\phi} + \frac{3H\dot{\phi}}{2\beta - 1} + \left(\frac{V'}{\beta(2\beta - 1)}\right)\left(\frac{2M^4}{\dot{\phi}^2}\right)^{\beta - 1} = 0, 
\end{eqnarray}
which can also be obtained from the total energy conservation law ($u_{\mu}T^{\mu\nu}_{;\nu}=0$).

Let us now prove that a canonical scalar field cannot describe both the smooth dynamic and thermodynamics of a running $\Lambda$(H)-term (from de Sitter to de Sitter), as discussed in the previous sections. 

\subsection{The Failure of the Canonical Case}

The generic non-canonical Lagrangian (\ref{L1}) can also be seen as non-linear extension of the ordinary scalar field Lagrangian, which is readily recovered by taking $\beta=1$ in (\ref{L1}). One finds: 
\begin{equation}
\mathcal{L}(X,\phi) \equiv X - V(\phi) = \frac {1}{2} {\dot \phi}^2 - V(\phi).
\end{equation}
As should be expected, the energy density (\ref{ED}) and pressure (\ref{p}) of the field now reduces to:

\begin{eqnarray}
\rho_{\phi} & \equiv & \rho_k + V(\phi)=\frac{{\dot\phi}^2}{2} + V({\phi}),\\ 
p_{\phi} & \equiv & p_k - V (\phi) = \frac{{\dot\phi}^2}{2} - V(\phi), 
\end{eqnarray}
where $\rho_k$ and $p_k$ are, respectively, the kinetic energy density and pressure of the ordinary canonical scalar field.
 
On the other hand, it is also widely known that such a scalar field can be interpreted as a mixture of {\it two interacting perfect fluids} with different equations of state, namely: A material Zeldovich's stiff fluid \cite{Zeld62}, where ($\rho_k=p_k ={{\dot\phi}^2}/{2}$), plus a pure vacuum ($\Lambda$-term) obeying an EoS $\rho_v=V(\phi)=-p_v$ (see e.g., \cite{C2002} for a more detailed discussion with examples). Therefore, as happens in the running vacuum model, by assuming that both components are gravitationally coupled, in principle, there will be no local energy-momentum conservation for each perfect fluid component separately. Only the total energy-momentum tensor of the system as a whole is conserved. 

Nevertheless, under such an interpretation, whether the potential of the unstable vacuum dominates (maximum scale $H_I$) and decays spontaneously through a non-adiabatic decaying process conserving the specific entropy (as discussed in section 3), the energy stored in the field will be transferred to a stiff component. In this way, the standard thermal bath formed by ultra-relativistic particles ($p_k=\rho_k/3$) is not generated. An inevitable consequence is that a new phase transition needs to be hypothesised (perhaps through a new coupling term) thereby forcing the formation of a thermal bath with the universe finally entering in the standard radiation phase. Naturally, in this case, the two fluid description without additional assumption is unable to describe faithfully the complete scenario driven by the running vacuum cosmology. 
 
Such a scenario will happen naturally when  coherent field oscillation phase is absent with the vacuum field ($\Lambda$ term) approaching continuously its final, very low (but finite!), which is defined by the scale $H_F$. As we shall see next, such a picture allow us to represent an evolution driven by the same scalar field (from de Sitter to de Sitter) based on a non-canonical scalar field and mimicking the running vacuum term as discussed in previous sections. 
 
\subsection{The Non-Canonical Case}
Let us now consider the generic non-standard case. To begin with we remark from equations (\ref{ED}) and (\ref{p}) that the general EoS parameter for the kinetic term (the material fluid) can be defined as:
\begin{equation}\label{EoS1}
\omega \equiv \frac{p_k}{\rho_k} = \frac{1}{2\beta -1}.
\end{equation}
Therefore, based on the two-fluid interpretation, the kinetic term for $\beta = 2$ obeys exactly the radiation EoS \footnote{This kind of single-fluid EoS can also be obtained from a k-essence field with Lagrangian $L(g(\phi),X) = g(\phi)F(X)$ \cite{SRJ23}. However, as remarked in \cite{FM10}, a two fluid description with dissipation is possible only if another interacting k-essence is considered.}. Thus, by assuming as before that the classical universe emerged from the quantum gravity regime as a de Sitter like solution with scale $H_I$ (see {\bf Figure 1}), this means that $\rho_k (H_I)=p_k(H_I)=0$, and the unstable potential $V(\phi)$ must decay directly in ultra-relativistic particles. Interestingly, the inflationary process is not adiabatic since the primeval thermal bath is a consequence of the energy transfer from the potential $V(\phi (H))$ or equivalently, the $\Lambda$(H)-term to the radiation created component (see section III). In certain sense, this inflationary scenario resembles some variants of the warm inflationary process because the primeval accelerating expansion is not adiabatic. However, it is somewhat different because the whole thermal bath here is a consequence of the running vacuum process (de Sitter instability) and not related with an initial singular state as assumed in some warm inflationary scenarios \cite{BF1995,AB1995,AB1995_2,ML2000}. 

Now let us demonstrate that the general equation of motion for the running $\Lambda(H)$ cosmology (\ref{eqmf}) is reproduced by the non-canonical scalar field based on the two

\begin{figure}[!ht]
    \centering
    \includegraphics[scale = 0.6]{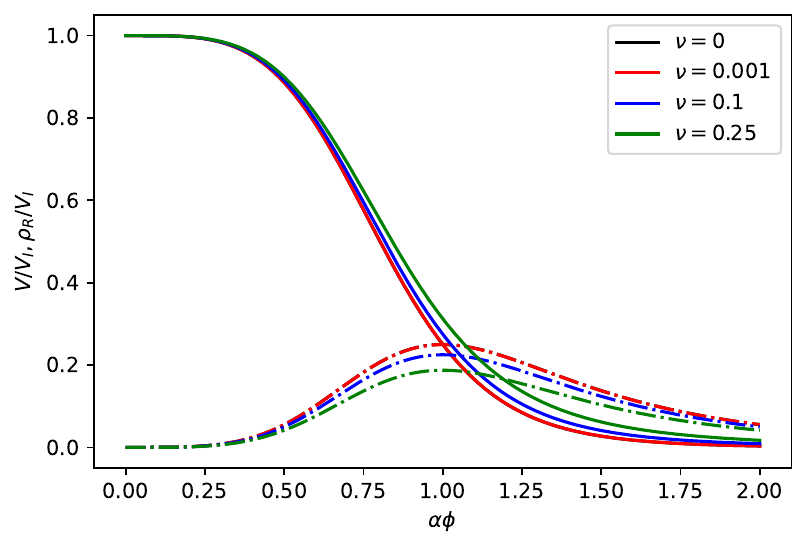}
    \caption{Dimensionless potential and kinetic energy (radiation) density. The solid lines describe the evolution of the potential with respect to $\alpha\phi$ and some selected values of the parameter $\nu$ depicted in the inner rectangle. Dashed lines display the same behaviour to the radiation energy density. In the beginning of the classical evolution we have only an emergent de Sitter universe for all values of $\nu$\, ($V=V_I, \rho_{R}=0$). Notice that all the created particles are the leftover of the primeval de Sitter vacuum state ($V=V_I$) for $H=H_I$ and $\phi=0$. For $\phi =\alpha^{-1}$, $\rho_R = V(\alpha\phi)$, $H=H_e$ and inflation ends [see also (\ref{He}) and Figure 1].} 
    \label{rad_pot}
\end{figure}

\noindent fluid interpretation as above proposed. In this case, the {\bf EFE} equations (\ref{eq1}) and (\ref{eq2}), now take the form:

\begin{eqnarray}
\rho_k\,+ V(\phi) &=& 3M_{P}^2\,H^2\;,\label{E1a}\\
p_k - V(\phi) &=& -M_{P}^2\left[2\dot H + 3H^{2}\right]\,,\label{E2a}
\end{eqnarray}
where $\rho_k$ and $p_k$ are the kinetic energy density and pressure of the fluid satisfying the EoS (\ref{EoS1}) while $V(\phi)$ must be determined in such a way that $\Lambda (H(\phi))= M_P^{-2}V(\phi)$. 
Interestingly, the above equations imply that the cosmic dynamics is also driven by: 
\begin{equation}\label{Eqm2}
 \dot H + \frac{3(1+\omega)}{2}H^2 \left[1 - \Omega_{\Lambda}(H)\right]=0\,, 
\end{equation}
where the equation of state $p=\omega \rho$ $(\omega \geq 0)$ was used and $\Omega_{\Lambda} (H)=V(H(\phi))/3M_{P}^2H^{2}$ is the vacuum density parameter. 

\begin{figure}[!ht]
    \centering
    \includegraphics[scale = 0.6]{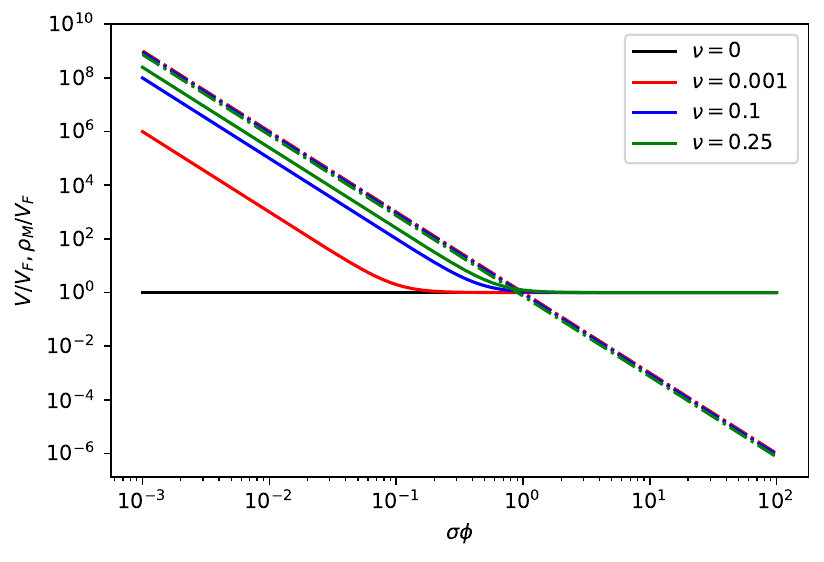}
    \caption{Dimensionless potential and matter energy density. As in the Figure 2, different solid lines describe the evolution of the potential with respect to $\sigma\phi$, for some selected values of the parameter $\nu$ depicted in inner rectangle. The dashed lines display the same behaviour for the dimensionless matter energy density. Notice that for all allowed values of $\nu$, the model evolves in the long run to a de Sitter spacetime described by the  scale $H_F$ [see also (\ref{V_matter}) and Figure 1].}
    
\end{figure}

Once the two-fluid interpretation  is assumed,  the 'H-equation of motion' is independent of the particular expression of the potential. Nevertheless, an expression for $V(\phi)$ need to be given and the scalar field solution properly derived. 

Inspired by the derived expression (\ref{LH}) for $\Lambda(H)$, let us discuss the potential contributions emulating the different cosmic phases starting from the primeval de Sitter state. 

\subsection{From de Sitter to the Radiation-Vacuum Dominated Era}
To begin with, let's consider $\beta=2$. The energy density and pressure of the non-canonical scalar field becomes:

\begin{eqnarray}
\rho_{\phi} &\equiv& \rho_k + V(\phi) = \frac{3\dot{\phi}^4}{4M^4} + V(\phi), \label{rhok} \\
p_{\phi} &\equiv& \frac{1}{3} \rho_k - V(\phi) = \frac{\dot{\phi}^4}{4M^4} - V(\phi), \label{pek}
\end{eqnarray}
with the kinetic component emulating  the radiation EoS. Now, in order to describe the initial de Sitter configuration and its transition to a radiation dominated phase, let us also  assume the following scalar field potential: 

\begin{equation}\label{Pot}
V(\phi) = V_I \frac{1 + \nu(\alpha\phi)^{4}}{[1 + (\alpha\phi)^{4}]^{2}},\,\,\, \phi \geq 0.
\end{equation} 
where   $V_I = V(\phi=0) =3M_P^{2}{H^{2}_I}$ and the constant $\alpha$ is a dimensional parameter, [$\alpha] = [\phi]^{-1}$. Its value is calculated in the Appendix C  
 \begin{equation}
  \alpha = \frac{\left[(1 - 2\nu)(1 - \nu)^3\right]^{1/4}}{M}\sqrt{\frac{H_I}{2M_P}}. 
 \end{equation}


In order to obtain the expression for the radiation energy density let us assume the \textit{ansatz} $\rho_k=(1 - \nu)(\alpha\phi)^{4}V(\phi)/[1 + \nu(\alpha\phi)^{4}]$. Inserting this into (\ref{rhok}) we obtain:
\begin{equation}\label{rhor_r_phi}
  \rho_k(\phi) = V_I\frac{(1 - \nu)(\alpha\phi)^4}{[1 + (\alpha\phi)^4]^2},
\end{equation}
and
\begin{equation}\label{H_r_phi}
  H(\phi) = \frac{1}{(1 - \nu)}\frac{\dot{\phi}}{\phi} = \frac{H_I}{[1 + (\alpha\phi)^4]^{1/2}}.
\end{equation}

Now, by inserting the expression for $H(\phi)$ above in equations (\ref{Pot}) and (\ref{rhor_r_phi}), one can show that:

\begin{eqnarray}
  V(H(\phi)) &=& 3\nu M_P^2H^2 + 3(1 - \nu)M_P^2H^2\left(\frac{H}{H_I}\right)^2, \\
  \rho_k &=& 3(1 - \nu) M_P^2H^2\left[1 - \left(\frac{H}{H_I}\right)^2\right],
\end{eqnarray}
thereby recovering the expressions (\ref{rhoL1}) and (\ref{rhor}) shown in section IV. Hence,  the proposed potential (\ref{Pot}) emulates completely the cosmological dynamics of the vacuum-radiation dominated era (\ref{eqmf1}). In particular, we see that there is no a thermal bath for $H=H_I$, since in this case $\rho_k(H_I)=0$.

In {\bf Figure 2} we display the evolution of the basic dimensionless quantities, namely, the potential scalar field and  the kinetic term $(\rho_k)$. The latter one describes the created component forming the thermal bath of effectively massless particles ($\omega=1/3$). As one may check, the transition from vacuum to the radiation-vacuum phase is also determined by the condition $\rho_k=V(\phi)$ [see discussion below (\ref{dec_par})].  

\subsection{From Matter-Vacuum to the final de Sitter Era}

In this case $\omega=0$ which means that $\beta >> 1$ in (\ref{EoS1}). Now, based on the expression (\ref{LH}) for $\Lambda(H)$, we propose the following potential $V(\phi(H))$ to the matter dominated era:
\begin{equation}\label{V_matter}
  V(\phi) = V_F\left[1 + \nu(\sigma\phi)^{-3}\right], 
\end{equation}
where $V(\phi)=3Mp^{2}H^2$ and $\sigma$  as before is also a dimensional parameter, $[\sigma]\equiv[\phi]^{-1}$. As a function of the scales $H_F$ and $M$ and the parameter $\nu$, it is given by (see Appendix C)
\begin{equation}
 \sigma = (1 - \nu)\frac{H_F}{M^{2}}.   
\end{equation}

In order to obtain the expression for the matter energy density we will again consider a similar \textit{ansatz} of the transition from de Sitter ($H_I$) to the radiation-vacuum dominated phase, namely: $\rho_k=(1 - \nu)(\sigma\phi)^{-3}V(\phi)/[1 + \nu (\sigma\phi)^{-3}]$. Now, keeping this in mind it is easy to show that:

\begin{equation}\label{rhor_M_Phi}
  \rho_k(\phi) = (1 - \nu)V_F(\sigma\phi)^{-3}
\end{equation}
and
\begin{equation}\label{H_M_Phi}
  H(\phi) = \frac{1}{(1 - \nu)}\frac{\dot{\phi}}{\phi} = H_F\left[1 + (\sigma\phi)^{-3}\right]^{1/2}.
\end{equation}

By substituting the expression for $H(\phi)$ above in both equations (\ref{V_matter}) and (\ref{rhor_M_Phi}), one can show that:

\begin{eqnarray}
  V(\phi(H)) &=& 3(1 - \nu)M_P^2H_F^2 + 3\nu M_P^2H^2,
  \\
  \rho_k &=& 3(1 - \nu) M_P^2H^2\left[1 - \left(\frac{H_F}{H}\right)^2\right],
\end{eqnarray}
\noindent recovering the expressions (\ref{rhoL}) and (\ref{rhoM}) shown in section IV. As it happens for the early universe, the potential (\ref{V_matter}) recovers the cosmological dynamics of the matter-vacuum era (\ref{eqmf2}).

\vskip 0.5cm

\section{Final Remarks and Conclusions} Entropy is the quantity which allows us to describe the progress of non-equilibrium dissipative processes. A special ingredient of our scenario discussed here is that it takes into account the second law of thermodynamics from the very beginning. For instance, the non-equilibrium thermodynamics was explicitly used to establish an integral relation uniting  $\Lambda(H)$ and the particle creation rate $\Gamma(H)$. The running vacuum emerges from an unknown quantum gravity regime  as a primeval de Sitter vacuum. Due to a structural thermal instability, it transfers irreversibly energy, particles and entropy to the spacetime forming the primeval thermal bath at the expenses of its own energy density at scale, which is initially defined by $H_I$, which is smaller than the reduced Planck mass. In this connection see also \cite{LA2021} where the dynamic evolution of this phase forming the thermal bath was investigated for $\nu=0$).

Subsequently, the evolution departs slightly from the standard radiation and matter dominated phases sustained by a softer generation of entropy. Since the scale $H_F$ is incredibly low, for all practical purposes, the final vacuum state will be eternal. 

There are also two interesting aspects of the scenario proposed here. Firstly, the irreversible evolution of the running vacuum generating a complete cosmology, can also be described by a single minimally coupled non-canonical scalar field when it is interpreted as a mixture of two interacting perfect fluids, as suggested long ago (see section V). Secondly,  all the energy scales appearing in the potential and also in the non-canonical kinetic term are sub-Planckian with no fine-tuning.  

Finally, we stress that the mild deviation from the rigid (i.e., non-dynamical) vacuum model ($\Lambda$CDM), may also provide an explanation to the current $H_0$ and $\sigma_8$ tensions, perhaps simpler than other phenomenological dark energy interacting models. Studies along these lines are in progress and, naturally, still deserves a closer scrutiny.

\vspace{0.3cm}

\noindent {\bf Acknowledgments:} P. E. M. Almeida is grateful to a PhD fellowship from  CAPES and J. A. S. Lima is partially supported by CNPq under grant 310038/2019-7, CAPES (88881.068485/2014), and FAPESP (LLAMA Project No. 11/51676-9). 
\appendix

\section{``Adiabatic" Creation ({$\dot \sigma = 0$}) and some consequences}

 In this appendix we demonstrate that the specific entropy (per particle) in our approach remains constant ({$\dot \sigma = 0$}), and also the validity of equation (\ref{dotS}). 

The specific entropy is defined by

\begin{equation}
    \sigma \equiv \frac{s}{n} \Rightarrow \dot{\sigma} = \sigma\left(\frac{\dot{s}}{s} - \frac{\dot{n}}{n}\right).
\end{equation}

From equations (\ref{n_gamma}) and (\ref{s_gamma}) one may check that

\begin{equation}\label{A2}
    \frac{\dot{n}}{n} = \frac{\dot{s}}{s} = \Gamma - 3H,
\end{equation}
and, therefore, $\dot{\sigma} = 0$.

Finally, by defining the expressions for the comoving number of particles, $N=na^{3}$, and entropy of the fluid, $S=sa^{3}$, it is straightforward to compute the time derivatives:
\begin{eqnarray}
    \frac{\dot{N}}{N} = \frac{1}{na^3}\frac{d(na^3)}{dt} &=& \frac{\dot{n}}{n} + 3H = \Gamma,
\\
    \frac{\dot{S}}{S} = \frac{1}{sa^3}\frac{d(sa^3)}{dt} &=& \frac{\dot{s}}{s} + 3H = \Gamma,
\end{eqnarray}
where the last equality in both equations comes from  (\ref{A2}). Both results were summarised in (\ref{dotS}).
Note that although depending on the symmetries of the FLRW geometry, equation (\ref{dotS}) is independent of Einstein's field equations.

\section{Determination of $\gamma$ parameter and the constant $B$}
Let us now consider a quadratic correction in the creation rate $\Gamma(H)=3\nu H$ (see section III):

\begin{equation}\label{gammaF}
\Gamma(H) = 3\nu H \left[1 +  \frac{\gamma}{\nu}\left({H}/{H_I}\right)^{2}\right],  
\end{equation}
where the numerical factors are for mathematical convenience. In this case,  a simple integration  of  (\ref{Leq}) provides an expression for $\Lambda(H)$  in terms of the constant $B$ and also the set of free parameters ($H_I,\nu$) appearing above. We find 
\begin{equation}\label{LH_2}
\Lambda(H) = B + 3\nu H^{2} + \frac{3\gamma{H^{2}}}{2}\left(\frac{H}{H_I}\right)^{2}.
\end{equation}
At low redshifts ($H<<H_I$), the last term on the {\it r.h.s.} is negligible and by using the late time boundary condition $\Lambda(H_F) = 3H_F^{2}$, a value $B= 3(1-\nu)H_F^{2}$ is obtained. In the opposite extreme,  $H>>H_F$, the constant $B$ is negligible and using de Sitter condition at early times, $\Lambda(H_I) = 3H_I^{2}$, it follows that $\gamma=2(1-\nu)$. Finally, by inserting such values, the result (\ref{LH}) is recovered.

\section{Determination of $\alpha$ and $\sigma$ parameters}
In this appendix we determine the values of $\alpha$ and $\sigma$ parameters appearing in the expressions of the scalar field potential to the vacuum-radiation and matter-vacuum phases. To begin with, we first consider that part of the derivation which is common for both quantities.  From Einstein field equations  (\ref{E1a}) and (\ref{E2a}), the  kinetic term of the non-canonical scalar field energy density can be written as:
\begin{equation}\label{A1}
  \rho_k = -\frac{2M_{P}^2\dot{H}}{(1 + \omega)},
\end{equation}
where the equality is also a consequence of the kinetic equation of state (\ref{EoS1}). Let us now consider each case  separately.\\

{\bf(i)} $\alpha$ parameter. By taking $\omega=1/3$ and using the definition of $\rho_k$ from (\ref{rhok}), the above equation implies that $\dot{\phi}$ can be written as: 

\begin{equation}
    \dot{\phi}^4 = -2M^4M_{P}^2\dot{H}, 
\end{equation}
and replacing $\dot{H}$ from the cosmological `H-equation of motion'  (\ref{Eqm2}) we obtain: 
\begin{eqnarray}\label{dotphi}
    \dot{\phi} = \sqrt{2}(1 - \nu)^{1/4}MM_{P}^{1/2}H^{1/2}\left[1 - \frac{H^2}{H_I^2}\right]^{1/4},
\end{eqnarray}
where we have discarded $\dot{\phi} < 0$. Actually,  $\phi \geq 0$ and its value is always increasing. Now, remembering that $H = \dot{\phi}/[(1 - \nu)\phi]$ ({\bf \ref{H_r_phi}}) the above equation can be rewritten as: 

\begin{equation}\label{phi_alpha}
    \phi = \frac{\sqrt{2}MM_{P}^{1/2}}{(1 - \nu)^{3/4}H^{1/2}}\left[1 - \frac{H^2}{H_I^2}\right]^{1/4}.
\end{equation}
Hence, recalling that $\phi_e = \alpha^{-1}$ and that at the end of inflation the Hubble parameter is $H(\phi_e) = H_I\sqrt{1 - 2\nu/2(1 - \nu)}$, the above equation at $\phi = \phi_e$ becomes: 

\begin{eqnarray}
    \phi_e &=& \alpha^{-1} = \frac{\sqrt{2}MM_{P}^{1/2}}{(1 - \nu)^{3/4}H_e^{1/2}}\left[1 - \frac{H_e^2}{H_I^2}\right]^{1/4} 
    \\
    &=& \sqrt{2}\left[\frac{1}{(1 - \nu)^3(1 - 2\nu)}\right]^{1/4}M\sqrt{\frac{M_P}{H_I}}. 
\end{eqnarray}
so that the  expression for $\alpha$ is given by:

\begin{equation}
    \alpha = \frac{1}{M}\left[(1 - \nu)^3(1 - 2\nu)\right]^{1/4}\sqrt{\frac{H_I}{2M_P}}.
\end{equation}
Note that beyond the reduced Planck mass $\alpha$ depends on the relevant free scales of the two vacuum-radiation phase ($H_I, M$). This means that a large class of sub-Planckian values are accessible even assuming $H_I =M_p$. \\

 ({\bf ii}) $\sigma$ parameter: In this case $\beta >>1$ and $\omega=0$ while the expression of $V(\phi)$ in the matter phase is given by (\ref{V_matter}). Now, following the same  procedure of the previous case, we see from  (\ref{A1}) that
 \begin{equation}
 \rho_k = -2M_{P}^2\dot{H}\,\,\, \Rightarrow \,\,\,\dot{\phi}^{2\beta} = -\frac{2^{(1 + \beta)}M^{4(\beta - 1)}M_{P}^2\dot{H}}{2\beta - 1}. 
\end{equation}

Let us rewrite $\dot{\phi}$ using the fact that $2\beta - 1 = 1/\omega$ (\ref{EoS1}):

\begin{equation}
    \dot{\phi}^{\frac{1 + \omega}{\omega}} = -\omega2^{\frac{1 + 3\omega}{2\omega}}M_P^2 M^{4\left(\frac{1 -\omega}{2\omega}\right)}\dot{H}.
\end{equation}

Substituting the cosmological `H-equation of motion' for the matter dominated phase (\ref{eqmf2}) in the expression above, one can write

\begin{equation}
   \dot{\phi}^{\frac{1 + \omega}{\omega}} = 3\omega(1 - \nu)2^{\frac{1 + \omega}{2\omega}}M_P^2 M^{4\left(\frac{1 -\omega}{2\omega}\right)}H^2\left[1 - \left(\frac{H_F}{H}\right)^2\right].
\end{equation}

In the same way we did for the determination of $\alpha$, we now utilise the relation between $H$ and  $\dot{\phi}$ and evaluate the resulting expression at $\phi = \sigma^{-1}$

\begin{equation}
    \sigma^{-\frac{1 + \omega}{\omega}} = \frac{3\omega.2^{\frac{1 + \omega}{2\omega}}M_P^2M^{4\left(\frac{1 -\omega}{2\omega}\right)}}{(1 - \nu)^{\frac{1}{\omega}}}H_{\sigma}^{\frac{\omega - 1}{\omega}}\left[1 - \left(\frac{H_F}{H_{\sigma}}\right)^2\right].
\end{equation}

From equation (\ref{H_M_Phi}) one can see that $H(\phi = \sigma^{-1}) \equiv H_{\sigma} = \sqrt{2}H_F$. Thus, the expression for $\sigma$ is given by

\begin{equation}
    \sigma = \frac{1}{(3\omega M_P^2)^{\frac{\omega}{1 + \omega}}}\left[\frac{1 - \nu}{M^{2(1 - \omega)H_F^{\omega - 1}}}\right]^{\frac{1}{1 + \omega}}.
\end{equation}
In the limit  $\omega \rightarrow 0$ we obtain:

\begin{equation}
    \sigma = (1 - \nu)\frac{H_F}{M^{2}},
\end{equation}
which differently from $\alpha$ does not depend on the initial de Sitter scale $H_I$.

\newpage{}

\bibliographystyle{ieeetrnew}
\bibliography{bib}

\begin{thebibliography}{10}

\bibitem{G1966}
E.~B. {Gliner}, ``{Algebraic Properties of the Energy-momentum Tensor and Vacuum-like States of Matter}'', {\em Soviet Journal of Experimental and Theoretical Physics}, vol.~22, p.~378, Feb. 1966.

\bibitem{S1966}
A.~D. {Sakharov}, ``{The Initial Stage of an Expanding Universe and the Appearance of a Nonuniform Distribution of Matter}'', {\em Soviet Journal of Experimental and Theoretical Physics}, vol.~22, p.~241, Jan. 1966.

\bibitem{M1973}
G.~L. Murphy, ``Big-bang model without singularities'', {\em Phys. Rev. D}, vol.~8, pp.~4231--4233, Dec 1973.

\bibitem{S80}
A.~A. {Starobinsky}, ``{A new type of isotropic cosmological models without singularity}'', {\em Physics Letters B}, vol.~91, pp.~99--102, Mar. 1980.

\bibitem{A83}
P.~Anderson, ``Effects of quantum fields on singularities and particle horizons in the early universe'', {\em Phys. Rev. D}, vol.~28, pp.~271--285, Jul 1983.

\bibitem{A84}
P.~R. Anderson, ``{Effects of quantum fields on singularities and particle horizons in the early universe. II}'', {\em Phys. Rev. D}, vol.~29, pp.~615--627, 1984.

\bibitem{Davies82}
N.~D. Birrell e P.~C.~W. Davies, {\em {Quantum Fields in Curved Space}}.
\newblock Cambridge Monographs on Mathematical Physics, Cambridge, UK: Cambridge Univ. Press, 2 1982.

\bibitem{AV1985}
A.~Vilenkin, ``Quantum origin of the universe'', {\em Nuclear Physics B}, vol.~252, pp.~141--152, 1985.

\bibitem{EM86}
E.~Mottola, ``Thermodynamic instability of de sitter space'', {\em Phys. Rev. D}, vol.~33, pp.~1616--1621, Mar 1986.

\bibitem{PRI89}
I.~Prigogine, J.~Geheniau, E.~Gunzig e P.~Nardone, ``Thermodynamics of cosmological matter creation'', {\em Proceedings of the National Academy of Sciences}, vol.~85, no.~20, pp.~7428--7432, 1988.

\bibitem{PRI89_2}
I.~{Prigogine}, J.~{Geheniau}, E.~{Gunzig} e P.~{Nardone}, ``{Thermodynamics and cosmology}'', {\em General Relativity and Gravitation}, vol.~21, pp.~767--776, Aug. 1989.

\bibitem{CLW92}
M.~O. Calvão, J.~A.~S. Lima e I.~Waga, ``On the thermodynamics of matter creation in cosmology'', {\em Physics Letters A}, vol.~162, no.~3, pp.~223--226, 1992.

\bibitem{CLW92_2}
J.~A.~S. Lima, M.~O. Calvao e I.~Waga, ``Cosmology, thermodynamics and matter creation'', 2007.

\bibitem{LG92}
J.~A.~S. Lima e A.~Germano, ``On the equivalence of bulk viscosity and matter creation'', {\em Physics Letters A}, vol.~170, no.~5, pp.~373--378, 1992.

\bibitem{SW89}
S.~Weinberg, ``The cosmological constant problem'', {\em Rev. Mod. Phys.}, vol.~61, pp.~1--23, Jan 1989.

\bibitem{V2014}
H.~E.~S. Velten, R.~F. vom Marttens e W.~Zimdahl, ``Aspects of the cosmological “coincidence problem”'', {\em The European Physical Journal C}, vol.~74, Nov. 2014.

\bibitem{R2019}
A.~G. Riess, S.~Casertano, W.~Yuan, L.~M. Macri e D.~Scolnic, ``Large magellanic cloud cepheid standards provide a 1\% foundation for the determination of the hubble constant and stronger evidence for physics beyond {$\Lambda$CDM}'', {\em The Astrophysical Journal}, vol.~876, May 2019.

\bibitem{LCM2007}
J.~V. Cunha, L.~Marassi e J.~A.~S. Lima, ``{Constraining H0 from the Sunyaev–Zel'dovich effect, galaxy cluster X-ray data and baryon oscillations}'', {\em Monthly Notices of the Royal Astronomical Society: Letters}, vol.~379, pp.~L1--L5, July 2007.

\bibitem{LCM2007_2}
J.~A.~S. {Lima} e J.~V. {Cunha}, ``{A 3\% Determination of H$_{0}$ at Intermediate Redshifts}'', {\em The Astrophysical Journal Letters}, vol.~781, p.~L38, Feb. 2014.

\bibitem{B2019}
S.~Birrer {\em et~al.}, ``{H0LiCOW – IX. Cosmographic analysis of the doubly imaged quasar SDSS 1206+4332 and a new measurement of the Hubble constant}'', {\em Monthly Notices of the Royal Astronomical Society}, vol.~484, pp.~4726--4753, 01 2019.

\bibitem{P2018}
N.~Aghanim {\em et~al.}, ``Planck2018 results: {VI}. cosmological parameters'', {\em Astronomy \&; Astrophysics}, vol.~641, p.~A6, Sept. 2020.

\bibitem{kids2020}
H.~Hildebrandt {\em et~al.}, ``{KiDS+VIKING-450:} cosmic shear tomography with optical and infrared data'', {\em Astronomy \&; Astrophysics}, vol.~633, p.~A69, Jan. 2020.

\bibitem{Val2021}
E.~{Di Valentino} {\em et~al.}, ``{Cosmology Intertwined III: f{\ensuremath{\sigma}}$_{8}$ and S$_{8}$}'', {\em Astroparticle Physics}, vol.~131, p.~102604, Sept. 2021.

\bibitem{ML2021}
M.~Lucca, ``{Multi-interacting dark energy and its cosmological implications}'', {\em Phys. Rev. D}, vol.~104, no.~8, p.~083510, 2021.

\bibitem{ST2009}
S.~Carneiro e R.~Tavakol, ``On vacuum density, the initial singularity and dark energy'', {\em General Relativity and Gravitation}, vol.~41, p.~2287, 05 2009.

\bibitem{LBS2013}
J.~A.~S. Lima, S.~Basilakos e J.~Solà, ``{Expansion history with decaying vacuum: a complete cosmological scenario}'', {\em Monthly Notices of the Royal Astronomical Society}, vol.~431, pp.~923--929, 03 2013.

\bibitem{PLBS2013}
E.~L.~D. Perico, J.~A.~S. Lima, S.~Basilakos e J.~Solà, ``Complete cosmic history with a dynamical {$\Lambda$(H)} term'', {\em Physical Review D}, vol.~88, Sept. 2013.

\bibitem{PLBS2013_2}
J.~A.~S. Lima, S.~Basilakos e J.~Solà, ``Nonsingular decaying vacuum cosmology and entropy production'', {\em General Relativity and Gravitation}, vol.~47, Mar. 2015.

\bibitem{J112015}
J.~Solà, A.~Gómez-Valent e J.~de~Cruz~Pérez, ``Hints of dynamical vacuum energy in the expanding universe'', {\em The Astrophysical Journal}, vol.~811, p.~L14, Sept. 2015.

\bibitem{GV2017}
A.~Gómez-Valent e J.~Solà, ``Relaxing the {$\sigma_8$} tension through running vacuum in the universe'', {\em EPL (Europhysics Letters)}, vol.~120, p.~39001, Nov. 2017.

\bibitem{S2017}
J.~Solà, A.~Gómez-Valent e J.~de~Cruz~Pérez, ``The {$H_0$} tension in light of vacuum dynamics in the universe'', {\em Physics Letters B}, vol.~774, p.~317–324, Nov. 2017.

\bibitem{ZSL2018}
G.~J.~M. Zilioti, R.~C. Santos e J.~A.~S. Lima, ``{From de Sitter to de Sitter: decaying vacuum models as a possible solution to the main cosmological problems}'', {\em Adv. High Energy Phys.}, vol.~2018, p.~6980486, 2018.

\bibitem{OT86}
M.~Özer e M.~O. Taha, ``A possible solution to the main cosmological problems'', {\em Physics Letters B}, vol.~171, no.~4, pp.~363--365, 1986.

\bibitem{OT86_2}
M.~Özer e M.~O. Taha, ``A model of the universe free of cosmological problems'', {\em Nuclear Physics B}, vol.~287, pp.~776--796, 1987.

\bibitem{KF1987}
K.~Freese, F.~C. Adams, J.~A. Frieman e E.~Mottola, ``Cosmology with decaying vacuum energy'', {\em Nuclear Physics B}, vol.~287, pp.~797--814, 1987.

\bibitem{CW1990}
W.~Chen e Y.-S. Wu, ``Implications of a cosmological constant varying as ${R}^{-2}$'', {\em Phys. Rev. D}, vol.~41, pp.~695--698, Jan 1990.

\bibitem{AR92}
A.-M.~M. Abdel-Rahman, ``Singularity-free decaying-vacuum cosmologies'', {\em Phys. Rev. D}, vol.~45, pp.~3497--3511, May 1992.

\bibitem{JCW92}
J.~C. Carvalho, J.~A.~S. Lima e I.~Waga, ``Cosmological consequences of a time-dependent $\ensuremath{\Lambda}$ term'', {\em Phys. Rev. D}, vol.~46, pp.~2404--2407, Sep 1992.

\bibitem{IW93}
I.~{Waga}, ``{Decaying Vacuum Flat Cosmological Models: Expressions for Some Observable Quantities and Their Properties}'', {\em \apj}, vol.~414, p.~436, Sept. 1993.

\bibitem{IW93_2}
V.~Silveira e I.~Waga, ``{Decaying Lambda cosmologies and power spectrum}'', {\em Phys. Rev. D}, vol.~50, pp.~4890--4894, 1994.

\bibitem{L96}
J.~A.~S. Lima, ``Thermodynamics of decaying vacuum cosmologies'', {\em Phys. Rev. D}, vol.~54, pp.~2571--2577, Aug 1996.

\bibitem{L96_2}
J.~A.~S. Lima, A.~I. Silva e S.~M. Viegas, ``{Is the radiation temperature-redshift relation of the standard cosmology in accordance with the data?}'', {\em Monthly Notices of the Royal Astronomical Society}, vol.~312, pp.~747--752, 03 2000.

\bibitem{LS2021}
J.~A.~S. Lima, S.~Trevisani e R.~Santos, ``Cosmic “adiabatic” photon creation: Temperature law and blackbody spectrum'', {\em Physics Letters B}, vol.~820, p.~136575, 2021.

\bibitem{LL86}
L.~D. Landau e E.~M. Lifshitz, {\em {Fluid Mechanics}}, vol.~6 of {\em Course of Theoretical Physics}.
\newblock Pergamon, second~ed., 1987.

\bibitem{LA2021}
J.~A.~S. Lima e P.~E.~M. Almeida, ``Noncanonical scalar fields and the primeval thermal bath'', {\em International Journal of Modern Physics D}, vol.~30, no.~14, p.~2142025, 2021.

\bibitem{SS2002}
I.~L. Shapiro e J.~Solà, ``The scaling evolution of the cosmological constant'', {\em Journal of High Energy Physics}, vol.~2002, p.~006–006, Feb. 2002.

\bibitem{BS2009}
S.~Basilakos, M.~Plionis e J.~Sol\`a, ``Hubble expansion and structure formation in time varying vacuum models'', {\em Phys. Rev. D}, vol.~80, p.~083511, Oct 2009.

\bibitem{KAM2016}
M.~Kamionkowski e E.~D. Kovetz, ``The quest for b modes from inflationary gravitational waves'', {\em Annual Review of Astronomy and Astrophysics}, vol.~54, p.~227–269, Sept. 2016.

\bibitem{VS2012}
S.~Unnikrishnan, V.~Sahni e A.~Toporensky, ``Refining inflation using non-canonical scalars'', {\em Journal of Cosmology and Astroparticle Physics}, vol.~2012, p.~018–018, Aug. 2012.

\bibitem{Zeld62}
Y.~B. Zel'dovich, ``{The equation of state at ultrahigh densities and its relativistic limitations}'', {\em Zh. Eksp. Teor. Fiz.}, vol.~41, pp.~1609--1615, 1961.

\bibitem{C2002}
L.~P. Chimento, ``Symmetry and inflation'', {\em Phys. Rev. D}, vol.~65, p.~063517, Mar 2002.

\bibitem{BF1995}
A.~Berera e L.-Z. Fang, ``Thermally induced density perturbations in the inflation era'', {\em Phys. Rev. Lett.}, vol.~74, pp.~1912--1915, Mar 1995.

\bibitem{AB1995}
A.~Berera, ``Warm inflation'', {\em Phys. Rev. Lett.}, vol.~75, pp.~3218--3221, Oct 1995.

\bibitem{AB1995_2}
A.~Berera, ``Thermal properties of an inflationary universe'', {\em Phys. Rev. D}, vol.~54, pp.~2519--2534, Aug 1996.

\bibitem{ML2000}
J.~M.~F. Maia e J.~A.~S. Lima, ``Extended warm inflation'', {\em Phys. Rev. D}, vol.~60, p.~101301, Oct 1999.

\bibitem{SRJ23}
J.~Socorro e J.~J. Rosales, ``Quantum fractionary cosmology: K-essence theory'', {\em Universe}, vol.~9, no.~4, 2023.

\bibitem{FM10}
F.~Arroja e M.~Sasaki, ``Note on the equivalence of a barotropic perfect fluid with a k-essence scalar field'', {\em Physical Review D}, vol.~81, May 2010.

\end{thebibliography}

\end{document}